\documentstyle[aps,12pt,preprint]{revtex}
\def \to {\rightarrow}
\def \beq {\begin{equation}}
\def \eeq {\end{equation}}
\def \ba {\begin{eqnarray}}
\def \ea {\end{eqnarray}}
\def \jpsi {J/\psi}
\def \ee {e^+e^-}
\def \< {\left <}
\def \> {\right >}

\begin{document}
\baselineskip 20pt
\renewcommand{\thesection}{\Roman{section}}
~
{\hfill PUTP-96-31}
\vskip 20mm
\begin{center}
{\Large \bf Determination of color-octet matrix elements from $\ee$ process
        at low energies}
\end{center}
\vskip 10mm
\centerline{Feng Yuan}
\vskip 2mm
\centerline{\small\it Department of Physics, Peking University, 
            Beijing 100871, People's Republic of China}
\vskip 4mm
\centerline{Cong-Feng Qiao}
\vskip 2mm
\centerline{\small\it China Center of Advanced Science and Technology
(World Laboratory), Beijing 100080, People's Republic of China}
\vskip 4mm
\centerline{Kuang-Ta Chao}
\vskip 2mm
\centerline{\small\it China Center of Advanced Science and Technology
(World Laboratory), Beijing 100080, People's Republic of China}
\vskip 1mm
\centerline{\small\it Department of Physics, Peking University, 
            Beijing 100871, People's Republic of China}
\vskip 15mm

\begin{center}
{\bf\large Abstract}

\begin{minipage}{140mm}
\vskip 5mm
\par

We present an analysis of the preliminary experimental data of direct $\jpsi$
production in $\ee$ process at low energies. 
We find that the color-octet contributions are crucially important to the 
cross section at this energy region, and their inclusion produces a
good description of the data.
By fitting to the data, we extract the individual values of two color-octet 
matrix elements:
$\< {\cal O}_8^{\psi}(^1S_0)\> \approx 1.1\times 10^{-2} GeV^3$,
$\< {\cal O}_8^{\psi}(^3P_0)\> /m_c^2\approx 7.4\times 10^{-3}GeV^3$.
We discuss the allowed range of the two matrix elements constrained by 
the theoretical uncertainties.
We find that $\< {\cal O}_8^{\psi}(^1S_0)\> $ is poorly determined because it
is sensitive to the variation of the choice of $m_c$, $\alpha_s$ and
$\< {\cal O}_1^{\psi}(^3S_1)\> $.
However $\< {\cal O}_8^{\psi}(^3P_0)\> /m_c^2$ is quite stable
(about $(6-9)\times 10^{-3}GeV^3$) when the parameters vary in reasonable
ranges.
The uncertainties due to large experimental errors are also discussed.

\vskip 5mm
\noindent
PACS number(s): 13.60.Le, 13.20.Gd
\end{minipage}
\end{center}
\vfill\eject\pagestyle{plain}\setcounter{page}{1}

In recent years, NRQCD factorization formalism\cite{bbl} has made impressive
progress in the study of heavy quarkonium production\cite{annrev}. In this
approach, the production process is factorized into short and long distance
parts, while the latter is associated with the nonperturbative matrix elements
of four fermion operators. The cross section for the inclusive production
of a quarkonium state $H$ can be expressed as a sum of products having the form
\beq
d\sigma(A+B\to H+X)=\sum\limits_n d\hat\sigma(A+B\to c\bar c[n]+X)
        \left <{\cal O}_n^H\right >.
\eeq
In the above, $d\hat\sigma$ represent the short distance coefficients associated
with the perturbative subprocesses in which a $c\bar c$ pair is produced in
a configuration denoted by $n$ (angular momentum $^{2S+1}L_J$ and color index 1 or 8).
$\left <{\cal O}_n^H\right >$ are the long distance nonperturbative 
matrix elements demonstrating the probability of a
$c\bar c$ pair evolving into the physical state $H$. $d\hat\sigma$ can
be obtained from perturbative calculations, while $\left <{\cal O}_n^H\right >$ are
nonperturbative parameters which can not be calculated perturbatively.

$\left <{\cal O}_n^H\right >$ consist of two kinds of matrix elements, {\it i.e.},
the color-singlet and color-octet matrix elements.
Color-singlet matrix elements may be related to the quarkonium radial wave
function or its derivatives at the origin, and may be calculated
by potential models or estimated by leptonic decay widths of quarkonium states.
Whereas, the nonperturbative color-octet matrix elements 
can only be determined from lattice QCD calculations or by fitting the
theoretical prediction of quarkonium production rates to the experimental data.
The preliminary lattice QCD calculations of quarkonium decay matrix elements
have been presented in the literature\cite{lqcd}.
For the quarkonium production matrix elements, before lattice QCD produces
its result, the NRQCD velocity scaling rules may be used to give a rough (order
of magnitude) estimate about size of the color-octet matrix elements. 
Under this velocity scaling rules, the matrix elements 
$\left <{\cal O}_n^H\right >$ (including both color-singlet and color-octet) are
related to each other by orders of $m_Qv^2$ or $v^2$, where $m_Q$ is the mass
of the heavy quark and $v$ is the typical relative velocity of the heavy
quark in the bound state.
For example, in $\jpsi$ production, the most important NRQCD matrix
elements are $\left <{\cal O}_1^\psi(^3S_1)\right >$, $\left <{\cal O}_8^\psi(^3S_1)\right >$,
$\left <{\cal O}_8^\psi(^1S_0)\right >$ and $\left <{\cal O}_8^\psi(^3P_J)\right >$, 
and according to the scaling rules, their relative sizes satisfy the following relations
\ba
\nonumber
  \< {\cal O}_1^\psi(^3S_1)\> \sim m_c^3v^3, &\< {\cal O}_8^\psi(^3S_1)\> \sim m_c^3v^7,\\
  \< {\cal O}_8^\psi(^1S_0)\> \sim m_c^3v^7, &\< {\cal O}_8^\psi(^3P_J)\> \sim m_c^5v^7.
\ea

Practically, the color-octet matrix elements have been determined by fitting 
the theoretical prediction of quarkonium production to the experimental data 
at various colliders.
According to the NRQCD factorization formalism, the matrix elements $\left <{\cal O}_n^H\right >$
are universal, so that their values measured from different colliders
must be the same. This is an important test to the color-octet production mechanism.
In previous studies\cite{braaten}\cite{cho1}, the matrix element $\< {\cal O}_8^\psi(^3S_1)\> $
is determined by fitting to the high $P_T$ prompt $\jpsi$
production at {\bf Fermilab Tevatron}, and their results are consistent
with the velocity scaling rules. However, the matrix elements 
$\< {\cal O}_8^\psi(^1S_0)\> $ and $\< {\cal O}_8^\psi(^3P_0)\> $ are not
determined individually, but as a linear combination\cite{cho1}.
Another linear combination of these two elements is also obtained in the
studies of $\jpsi$ photoproduction in $e^-p$ collisions\cite{photon} and
hadroproduction at fix target experiments\cite{fixed}, and their results are
not compatible with that of\cite{cho1}. Recently, some further investigations
on prompt $\jpsi$ production at the {\bf Tevatron} are performed\cite{benek}\cite{lhc}.
Their results show that the parton distributions strongly affect the extraction of the
values of the color-octet matrix elements, {\it i.e.}, different sets of parton
distributions result in different values of the matrix elements.
They also show that the initial- and final-state radiation are important in
prompt $\jpsi$ production at the {\bf Tevatron}\cite{lhc}.
All those progresses show that the inconsistency between the values measured
at {\bf CDF} and those measured in other processes may be 
mostly due to the large theoretical uncertainties in those calculations, such
as the parton distributions.

In contrast, the mechanism of $\jpsi$ production in $\ee$ annihilation process
is much clearer than those hadronic processes discussed above. The parton structure
is simpler, and there is no higher twist effects to be considered, so the 
theoretical uncertainty is much smaller.
Because in $\ee$ processes $\jpsi$ production has much
smaller theoretical uncertainty than in hadronic $\jpsi$ production
processes, 
it can be used to extract the color-octet matrix elements.
$\jpsi$ production at $\ee$ colliders has been investigated by several 
authors in literatures [10-17].
Braaten and Chen have noted that  
a clean signature of color-octet mechanism may be observed in the angular
distribution of $\jpsi$ production near the endpoint region at $\ee$ collider
in the low energy region such as at {\bf CLEO}\cite{chen}.
In this paper, we make use of the previous results of the calculations on the
$\jpsi$ production in $\ee$ process to extract the color-octet matrix elements
by comparing with the preliminary experimental data at low energies\cite{pluto}\cite{cleo}.

In our previous studies\cite{yuan1}, we have calculated color-singlet and color-octet
contributions to prompt $\jpsi$ production in $\ee$ annihilation at different 
energy scales under the NRQCD factorization formalism and
we have used the nonperturbative matrix elements which are consistent with the NRQCD
velocity scaling rules. We find that the color-octet contributions dominate over the color-singlet
contributions in all energy regions. At low energies, 
as pointed out in \cite{chen}\cite{yuan1}, the dominant production
channels are $\ee\to \jpsi +g$ via color-octet $^1S_0$ and $^3P_J$ processes,
while at high energies (say, above $20GeV$) the dominant channel is $\ee\to q\bar{q}g^*$ followed by $g^*$ fragmentation
to color-octet $\jpsi$ process.

For the color-octet processes
\beq
        \ee\to \gamma^*\to g+c\bar c[\b 8,^{2S+1}L_J],
\eeq
where $^{2S+1}L_J$ represent the $c\bar c$ pair states $^1S_0$ and $^3P_J$,
from Ref.\cite{chen}, we readily have 
\ba
\label{oct}
\sigma(\ee\to\jpsi g)&=&C_s \< {\cal O}_8^{\psi}(^1S_0)\> +C_p \< {\cal O}_8^{\psi}(^3P_0)\> ,
\ea
with
\ba
\label{cs}
C_s&=&\frac{64\pi^2e_c^2\alpha^2\alpha_s}{3}
        \frac{1-r}{s^2m},\\
\label{cp}
C_p&=&\frac{256\pi^2e_c^2\alpha^2\alpha_s}{9s^2m^3}
        \big [\frac{(1-3r)^2}{1-r}+\frac{6(1+r)}{1-r}+
        \frac{2(1+3r+6r^2)}{1-r}\big ],
\ea
where $r=m^2/s$, $m$ is the mass of $\jpsi$, and $s$ is the $\ee$ collision
c.m. energy squared.
Here we have used the approximate heavy quark spin symmetry relations
\beq
\< {\cal O}_8^{\psi}(^3P_J)\> \approx (2J+1)\< {\cal O}_8^{\psi}(^3P_0)\> .
\eeq

To extract the color-octet matrix elements, one needs to subtract the
color-singlet contributions from the total cross section.
The leading order color-singlet $\jpsi$ production rates at low energies
($<25 GeV$) comes from the process
\beq
        \ee\to \gamma^*\to \jpsi+gg.
\eeq
The cross section of this process is\cite{keung}
\beq
\frac{d\sigma(\ee\to\jpsi gg)}{\sigma_{\mu\mu}dzdx_1}=
\frac{64e_c^2 \alpha_s^2}{27}\frac{\< {\cal O}_1^{\psi}(^3S_1)\> }{m^3}
r^2 f(z,x_1;r),
\eeq
where 
\begin{eqnarray}
\label{e3}
\nonumber
f(z,x_1;r)&=&\frac{(2+x_2)x_2}{(2-z)^2(1-x_1-r)^2} +
 \frac{(2+x_1)x_1}{(2-z)^2(1-x_2-r)^2}\\ 
 \nonumber
 &+&
 \frac{(z-r)^2-1}{(1-x_2-r)^2(1-x_1-r)^2}
+ \frac{1}{(2-z)^2}\Big(\frac{6(1+r-z)^2}
 {(1-x_2-r)^2(1-x_1-r)^2}\\ 
& + & \frac{2(1-z)(1-r)}{(1-x_2-r)(1-x_1-r)r}
  +\frac{1}{r}\Big ),
\end{eqnarray}
and
$$\sigma_{\mu\mu}=\sigma_{QED}(\ee\to\mu^+\mu^-).$$
The variables $z,~x_i$ are defined as
\beq
z=\frac{2p\cdot k}{s},~~~x_i=\frac{2p_i\cdot k}{s},
\eeq
where $k$, $p$ and $p_i$ are the momenta of the virtual photon $\gamma^*$, $\jpsi$, 
and the outgoing gluon, respectively.

In the numerical calculations we use the following parameters as input
\beq 
\label{para}
m_c=1.5 GeV,~~~\alpha_s(2m_c)=0.26,~~~\< {\cal O}_1^{\psi}(^3S_1)\> =1.08 GeV^3\cite{lqcd}.
\eeq

In Fig.1, we show the theoretical prediction of direct $\jpsi$ production compared
with the experimental data (note that contribution from $\psi^\prime$ feeddown has been
subtracted from the data sample). The data at $\sqrt{s}=4\sim 5 GeV$ are from
{\bf PLUTO}\cite{pluto}, and those at $\sqrt{s}=10.6 GeV$ are from {\bf CLEOII}
\cite{cleo}\cite{kuhn}. The two color-octet matrix elements
$\< {\cal O}_8^{\psi}(^1S_0)\> $ and $\< {\cal O}_8^{\psi}(^3P_0)\> $ 
are treated as free parameters to be determined by fitting the data.
We use the data from two energy points to determine these two matrix elements.
One is from {\bf PLUTO} at $\sqrt{s}=5GeV$ and the other from {\bf CLEO} at
$\sqrt{s}=10.6GeV$. At the collision energies below $5GeV$ there might
exist some $c\bar c$ resonances which could contaminate the prompt $\jpsi$
production, that is why
we neglect the data in the energy region below $\sqrt{s}=5GeV$ to 
carry out the extraction.
And the extraction result is
\ba
\label{value}
\nonumber
\< {\cal O}_8^{\psi}(^1S_0)\> &=&1.1\times 10^{-2} GeV^3,\\
\frac{\< {\cal O}_8^{\psi}(^3P_0)\> }{m_c^2}&=&0.74\times 10^{-2} GeV^3,
\ea
where the central values of experimantal data ($R_\psi^{exp}=5.0\times 10^{-3}$
at $\sqrt{s}=5.0GeV$ and $R_\psi^{exp}=1.0\times 10^{-3}$ at $\sqrt{s}=10.6GeV$)
are used and the experimental errors are not included.
The dotted line in Fig.1 represents the contribution from color-octet $^1S_0$ 
subprocess, the dashed line is from color-octet $^3P_J$ process, and their 
sum is plotted as the dotted-dashed line, and the color-singlet contributions 
is shown as the short dashed line.
>From this figure, we can see clearly that the color-singlet contributions 
alone can not explain the observed cross sections in these energy regions.
After including the contributions from both color-singlet and color-octet
production processes (shown as the solid line), a satisfactory agreement 
between theoretical prediction and the experimental data will be achieved.
Especially, the cross section of direct $\jpsi$ production
at $\sqrt{s}=5GeV$ is much larger than that at $\sqrt{s}=10.6 GeV$. This
behavior can not be explained by the color-singlet model with perturbative
QCD. 

Here, we would like to emphasize that the {\bf PLUTO} data are due to direct
$\jpsi$ production but not the decays of $c\bar c$ resonances\cite{pluto}. Furthermore, we 
assume that in the energy region around $\sqrt{s}=5 GeV$, perturbative
QCD is approximately applicable to describe the $\jpsi$ production processes.
In this connection, we note that Driesen {\it et al.} have applied perturbative
QCD to calculate the $\jpsi$ production in $\ee$ process within the color-singlet
model\cite{kuhn}.
They could explain the experimental data from {\bf PLUTO} by using a large
running coupling constant, {\it i.e.},
setting $\alpha_s(Q^2)=1$ if $\sqrt{Q^2}<1GeV$.
Their results indicate that the main contribution to $\jpsi$ production in
this process comes from the region where the conjunctioned gluons
are soft.
It is not entirely clear whether this color-singlet $c\bar c$ plus a soft gluon can be factorized into
a higher Fock state for the physical $\jpsi$, such as color-octet ${}^3P_J$
states and color-octet ${}^1S_0$ state.
Nevertheless, if the experimental results of these two collaborations ({\bf CLEO} and
{\bf PLUTO}) are further confirmed, the color-octet production mechanism will 
probably provide a quite unique explanation.

In the following we estimate the theoretical uncertainties induced by the choice of the charm
quark mass, the strong coupling constant and the color-singlet matrix element
$\< {\cal O}_1^{\psi}(^3S_1)\> $.
First, we consider a variation of the charm quark
mass $m_c$ from $1.4GeV$ to $1.6GeV$ with other parameters unchanged (as
in (\ref{para})). This results in a variation of the
two color-octet matrix elements fitted values,
\ba
\label{value1}
\nonumber
\< {\cal O}_8^{\psi}(^1S_0)\> &=&-0.25~~{\rm to}~~ 2.3\times 10^{-2} GeV^3,\\
\frac{\< {\cal O}_8^{\psi}(^3P_0)\> }{m_c^2}&=&0.91~~{\rm to}~~0.61\times 10^{-2} GeV^3.
\ea
Also, a variation of the strong coupling constant $\alpha_s$ from $0.24$ to
$0.30$ will result in
\ba
\label{value2}
\nonumber
\< {\cal O}_8^{\psi}(^1S_0)\> &=&1.8~~{\rm to}~~ -0.009\times 10^{-2} GeV^3,\\
\frac{\< {\cal O}_8^{\psi}(^3P_0)\> }{m_c^2}&=&0.76~~{\rm to}~~ 0.71\times 10^{-2} GeV^3.
\ea
And a variation of the color-singlet matrix element 
$\< {\cal O}_1^{\psi}(^3S_1)\> $ from $1.0GeV^3$ to $1.2GeV^3$ will
result in
\ba
\label{value3}
\nonumber
\< {\cal O}_8^{\psi}(^1S_0)\> &=&1.4~~{\rm to}~~ 0.77\times 10^{-2} GeV^3,\\
\frac{\< {\cal O}_8^{\psi}(^3P_0)\> }{m_c^2}&=&0.72~~{\rm to} ~~0.76\times 10^{-2} GeV^3.
\ea
>From the above results, we can see that the fitted value of the element
$\< {\cal O}_8^{\psi}(^1S_0)\> $ is sensitive to a variation of the three
parameters $m_c$, $\alpha_s$ and $\< {\cal O}_8^{\psi}(^1S_0)\> $.
This is because, the cross section of $\jpsi$ production
is not sensitive to the value of the element $\< {\cal O}_8^{\psi}(^1S_0)\> $
({\it i.e.}, the coefficient of the $\< {\cal O}_8^{\psi}(^1S_0)\> $ term in the
expression of cross section is very small, see (\ref{oct}) (\ref{cs})).
In contrast, the fitted value of $\< {\cal O}_8^{\psi}(^3P_0)\> /m_c^2$
remains rather stable when the three parameters vary in reasonable ranges.
The above rough estimates may show 
the value of the element $\< {\cal O}_8^{\psi}(^3P_0)\> /m_c^2=(0.6-0.9)\times 10^{-2} GeV^3$.

Finally, we discuss how the present experimental errors can affect the fitted values
of the two matrix elements. 
Obviously, the large errors in experimental data will result in a larger error for the fitted values
of the two color-octet matrix elements. 
To see this, in  Fig.2 we show four curves for the $\jpsi$ production 
cross section, where the input parameters ($\alpha_s$, $m_c$, 
$\< {\cal O}_1^{\psi}(^3S_1)\> $) are same as in Fig.1. 
The two dotted lines correspond to
$\< {\cal O}_8^{\psi}(^3P_0)\> /m_c^2=7.4\times 10^{-3} GeV^3$,
$\< {\cal O}_8^{\psi}(^1S_0)\> =5.0\times 10^{-2} GeV^3$(up), and
$\< {\cal O}_8^{\psi}(^1S_0)\> =1.0\times 10^{-4}GeV^3$(down) respectively.
The two dashed lines correspond to 
$\< {\cal O}_8^{\psi}(^1S_0)\> =1.0\times 10^{-2} GeV^3$,
$\< {\cal O}_8^{\psi}(^3P_0)\> /m_c^2=2.0\times 10^{-2} GeV^3$(up), and
$\< {\cal O}_8^{\psi}(^3P_0)\> /m_c^2=2.0\times 10^{-3} GeV^3$(down) respectively.
We can see that the total cross section is not sensitive to
the value of $\< {\cal O}_8^{\psi}(^1S_0)\> $. 
When it increases from $1.0\times 10^{-4}GeV^3$ to $5.0\times 10^{-2}GeV^3$, 
the total cross section only changes a little.
But the value of $\< {\cal O}_8^{\psi}(^3P_0)\> $ varies rapidly as the data
value changes.
Therefore, in order to determine the color-octet matrix elements more 
rigorously, more precise measurement for the $\jpsi$ cross section at low
energies is apparently needed.

Meanwhile, we note that the color-octet matrix elements discussed in this
paper have also been determined from other experiments.
At the {\bf Tevatron}, prompt $\jpsi$ production at high $P_T$ has been
compared with theoretical predictions, and a global fit to all
$P_T$ region shows that at low $P_T$ the theoretical
prediction is dominated by the contributions from $<{\cal O}_8^{\psi}(^1S_0)>$ and $<{\cal O}_8^{\psi}(^3P_0)>$,
and the fitted result is\cite{cho1}
\beq
\label{teva}
\< {\cal O}_8^{\psi}(^1S_0)\>  +\frac{3}{m_c^2}\< {\cal O}_8^{\psi}(^3P_0)\> 
        =6.6\times 10^{-2}GeV^3.
\eeq
Recently, Beneke {\it et al.}\cite{benek} updated the extraction of color-octet
matrix elements by comparing the $P_T$ distribution for unpolarized direct
$\jpsi$ production with the most recent CDF data\cite{cdfrecent}.
By using the CTEQ4L PDF (parton distribution function) set (in their paper, they
have used three PDF sets, which resulted in different extractions\cite{benek}),
they obtained a different linear combination of the two color-octet matrix
elements,
\beq
\label{teva1}
\< {\cal O}_8^{\psi}(^1S_0)\>  +\frac{3.5}{m_c^2}\< {\cal O}_8^{\psi}(^3P_0)\> 
        =4.38\times 10^{-2}GeV^3.
\eeq
In Ref.\cite{lhc}, Sanchis-Lozano {\it et al.} used a Monte Carlo event
generator to treat the high order initial- and final-states radiation.
By fitting to the experimental data of prompt $\jpsi$ production at the {\bf Tevatron},
the authors obtained another extraction of these two color-octet matrix elements,
\beq
\label{teva2}
\< {\cal O}_8^{\psi}(^1S_0)\>  +\frac{3}{m_c^2}\< {\cal O}_8^{\psi}(^3P_0)\> 
        =1.44\times 10^{-2}GeV^3,
\eeq
where the CTEQ2L PDF set is used.
The value in Eq.(\ref{teva2}) is much smaller than those in Eq.(\ref{teva})
and (\ref{teva1}).
This shows that the inital-states radiation effects are crucialy important to
the extraction of these two color-octet matrix elements at the {\bf Tevatron}.

The studies of photoproduction at $e^-p$ collisions show that the
matrix element values in Eq.(\ref{teva}) may be overestimated\cite{photon}, 
and the authors obtained another linear combination of these two elements
from the forward $\jpsi$ photoproduction cross-section measurements,
\beq
\label{photo}
\< {\cal O}_8^{\psi}(^1S_0)\>  +\frac{7}{m_c^2}\< {\cal O}_8^{\psi}(^3P_0)\> 
        =2.0\times 10^{-2}GeV^3.
\eeq
The fixed-target hadroproduction result\cite{fixed} gives the same argument 
against the matrix elements values in Eq.(\ref{teva}), and gives
\beq
\label{fixed}
\< {\cal O}_8^{\psi}(^1S_0)\>  +\frac{7}{m_c^2}\< {\cal O}_8^{\psi}(^3P_0)\> 
        =3.0\times 10^{-2}GeV^3.
\eeq
Our results (\ref{value}) are not compatible with (\ref{photo}) and (\ref{fixed}).
The inconsistency may be mostly due to theoretical uncertainties and
large experimental errors.

However, it should be noted that there are large theoretical uncertainties
in our extraction of the color-octet matrix element
$\< {\cal O}_8^{\psi}(^1S_0)\> $, and also the large experimental errors may further
affect the extraction of the two matrix elements $\< {\cal O}_8^{\psi}(^1S_0)\> $
and $\< {\cal O}_8^{\psi}(^3P_0)\> $.
So, the compatibility of our results with those from {\bf Tevatron} and
the incompatibility of our results with those from fixed-target hadroproduction
and photoproduction should not be taken seriously.

In Ref.\cite{bdecay}, Fleming {\it et al.} make an analysis of the present extraction
values of the color-octet matrix elements.
After considering the constraints by the $\jpsi$ production rate in $b$-decays,
they obtained a rough range of the linear combination of the two color-octet
matrix elements,
\beq
\label{bdecay}
\< {\cal O}_8^{\psi}(^1S_0)\> +\frac{3}{m_c^2}\< {\cal O}_8^{\psi}(^3P_0)\> 
        =1.0\sim 6.0\times 10^{-2}GeV^3.
\eeq
This is compatible with all the above results (Eqs.(\ref{value}) and (\ref{teva})
through (\ref{fixed})). 

When this work is in progress, we receive a preprint of Fleming {\it et al.}\cite{ep},
in which they calculate the $\jpsi$ leptoproduction. They find that a negative 
value for the color-octet matrix element $\< {\cal O}_8^{\psi}(^3P_0)\> $ is 
still possible, which is inconsistent with our results. 
We would like to point out that if this matrix element is negative, the 
color-octet $^3P_J$ processes would provide a large negative contribution to 
$\jpsi$ production in $\ee$ process at low energies due to a large positive
coefficient for the matrix element $\< {\cal O}_8^{\psi}(^3P_0)\> $ at low 
energies (see Eqs.(4)-(6)). 
If so, the experimental data in Fig.1 can not be explained. The conflict 
between our results with those of \cite{ep} about the color-octet 
matrix elements puts forward a question to the universality of the color-octet 
matrix elements in these two situations.

In conclusion, in this paper we have calculated the direct $\jpsi$ production rates in
$\ee$ process at low energies. Our results show that the color-octet production
mechanism is significant to the production cross section in this energy region,
and this may provide another positive test for the validity of the color-octet
production mechanism. The extraction of elements $\< {\cal O}_8^{\psi}(^1S_0)\> $
and $\< {\cal O}_8^{\psi}(^3P_J)\> $ are performed by fitting to the
experimental data. But the large errors of the preliminary experimental
data can not give a precise estimate of the color-octet matrix elements.
We hope that more precise experimental data of direct $\jpsi$ production in
$\ee$ annihilation at {\bf CLEO II} and {\bf BEPC} will soon be obtained.

\vskip 1cm
\begin{center}
\bf\large\bf{Acknowledgements}
\end{center}

One of us (F.Yuan) thanks the staff of the Physics Department Computer Center (Room 540)
for their kind help.
This work was supported in part by the National Natural Science Foundation
of China, the State Education Commission of China and the State Commission
of Science and Technology of China.


\newpage
\centerline{\bf \large Figure Captions}
\vskip 2cm
\noindent
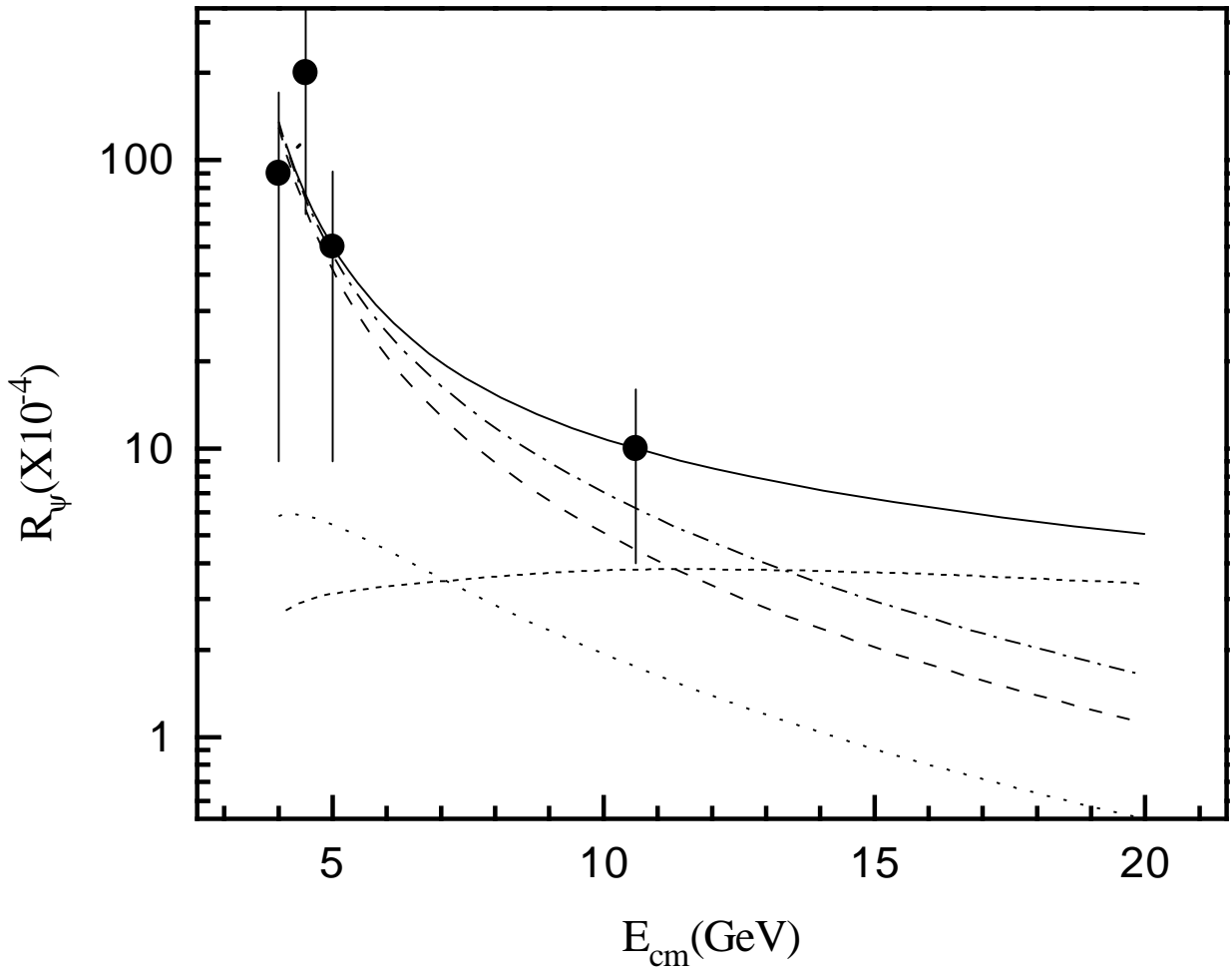
Fig.1. The total cross section of $\jpsi$ in $\ee$ process.
Color-octet $^1S_0$ process contribution is represented by the dotted line,
color-octet $^3P_J$ process by the dashed line, the sum of these two color-octet
processes by the dotted-dashed line, and color-singlet contribution by the short dashed
line. The experimental data are taken from Ref.\cite{cleo}\cite{pluto}.

\noindent
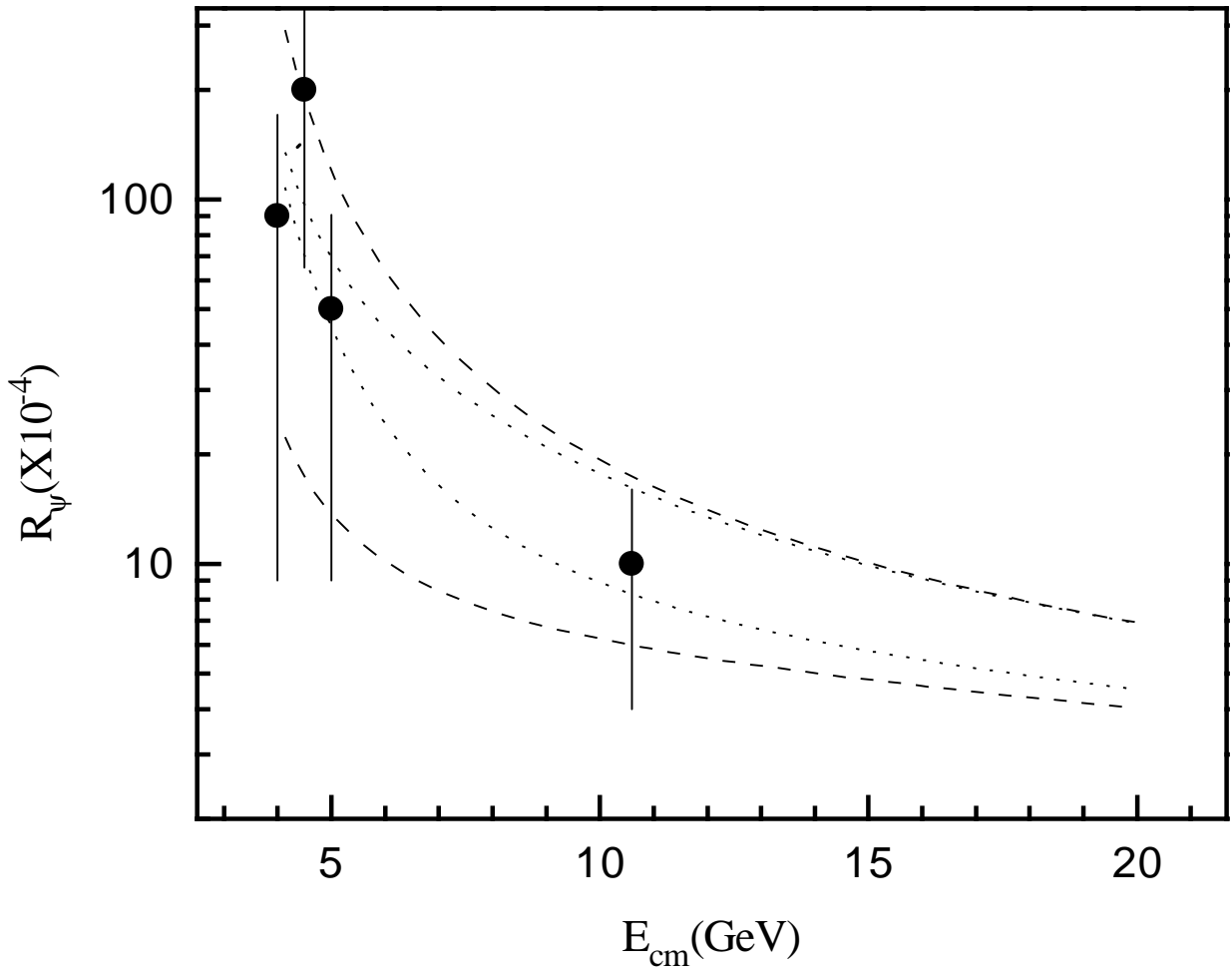
Fig.2. Description of the allowed values of the color-octet matrix elements
constrained by the experimental data.
The dotted lines correspond to
$\< {\cal O}_8^{\psi}(^3P_0)\> /m_c^2=7.4\times 10^{-3} GeV^3$,
$\< {\cal O}_8^{\psi}(^1S_0)\> =5.0\times 10^{-2} GeV^3$(up), and
$\< {\cal O}_8^{\psi}(^1S_0)\> =10^{-4}GeV^3$(down) respectively.
The dash lines correspond to
$\< {\cal O}_8^{\psi}(^1S_0)\> =1.0\times 10^{-2} GeV^3$,
$\< {\cal O}_8^{\psi}(^3P_0)\> /m_c^2=2.0\times 10^{-2} GeV^3$(up), and
$\< {\cal O}_8^{\psi}(^3P_0)\> /m_c^2=2.0\times 10^{-3} GeV^3$(down) respectively.
                       
\end{document}